\documentclass[12pt]{article}
\usepackage{amssymb}
\usepackage{amsmath}
\usepackage{amsfonts}
\usepackage{epsfig}
\usepackage{graphicx}

\begin{document}

\title{A consistent measure for lattice Yang-Mills}
\author{R. Vilela Mendes \thanks{%
Centro de Matem\'{a}tica e Aplica\c{c}\~{o}es Fundamentais, Universidade de
Lisboa, Edif\'{\i}cio C6, Campo Grande 1749-016 Lisboa (Portugal),
rvilela.mendes@gmail.com, rvmendes@fc.ul.pt} }
\date{ }
\maketitle

\begin{abstract}
The construction of a consistent measure for Yang-Mills is a precondition
for an accurate formulation of non-perturbative approaches to QCD, both
analytical and numerical. Using projective limits as subsets of Cartesian
products of homomorphisms from a lattice to the structure group, a
consistent interaction measure and an infinite-dimensional calculus has been
constructed for a theory of non-abelian generalized connections on a
hypercubic lattice. Here, after reviewing and clarifying past work, new
results are obtained for the mass gap when the structure group is compact.
\end{abstract}



\section{Introduction: Non-perturbative QCD and the Euclidean measure}

QCD, believed to be the theory of strong interactions, has the serious
shortcoming that only a limited sector can be treated analytically, namely
the one where short distance effects play a role. This is the perturbative
approach or, in the language of functional integration, the domain of
Gaussian measures. The perturbative approach to field theory is not entirely
satisfactory because perturbation theory is only an asymptotic expansion.
Nevertheless, in quantum electrodynamics, the coupling constant being small,
perturbation theory is applied with practical success. For QCD the coupling
is small only for distances which are much smaller than the radius of a
proton and therefore it is hopeless to use perturbative methods to calculate
light hadron masses, for example. Another aspect of QCD which is outside the
realm of perturbation theory is confinement, the fact that asymptotically we
do not observe the particles which correspond to the fundamental fields of
QCD, the quarks and gluons, while when the Compton wavelength of the
exchanged particles is very small the constituents of the hadrons behave
nearly like free particles. Since one cannot treat analytically distances of
the size of a hadron, confinement cannot be explained by perturbation theory.

To handle all these phenomena one has to rely on models or approximations,
the lattice approach, the introduction of local condensates, chiral symmetry
breaking models, the stochastic vacuum, etc. In all cases, in functional
integral terms, one needs to go beyond the Gaussian measure. Therefore a
precondition for the building of a reliable non-perturbative QCD, seems to
be the construction of a rigorous Euclidean measure for the Yang-Mills
theory. In particular a measure that handles in a consistent way both large
and small distance limits. The use of a consistent measure (in the sense to
be defined later) is also important for numerical calculations on the
lattice to insure that, when the lattice spacing approaches zero, one is
actually approaching the continuum limit.

First steps in the construction of such a measure were given in \cite%
{VilelaJMP} where a space for generalized connections was defined using
projective limits as subsets of Cartesian products of homomorphisms from
lattice based loops to a structure group. In this space, non-interacting and
interacting measures were defined as well as functions and operators. From
projective limits of test functions and distributions on products of compact
groups, a projective gauge triplet was obtained, which provides a framework
for an infinite-dimensional calculus in gauge theories. A central role is
played by the construction of an interacting measure which, satisfying a
consistency condition, can be extended to a projective limit of decreasing
lattice spacing and increasingly larger lattices.

Here the construction in \cite{VilelaJMP} is further clarified and extended
with a more detailed explanation on how the one-plaquette-at-a-time
refinement is performed in dimensions higher than two. An essential point in
this construction is the fact that the gauge covariant group elements that
one associates to each edge are loops based on an external point $x_{0}$.
Therefore several distinct independent loops may be associated to the same
edge.

In addition, some of the physical consequences of the constructed measure
will be explored, in particular the nature of the mass gap that it implies.
Here the main tool to be used is the theory of small random perturbations of
dynamical systems \cite{Freidlin2} \cite{Freidlin3}. This, as well as the
stochastic representation of the principal eigenvalue of elliptic equations
for arbitrary coupling intensities $g$, seems to be the most appropriate
tool to handle non-perturbative problems because its leading term is of
order $\exp \left( -\frac{C}{g^{2}}\right) $.

Although using rigorous mathematical concepts throughout the paper, the
constructions are kept as simple as possible, with the necessary
mathematical background explained both in the construction of the measure
and in the use of the theory of small random perturbations of dynamical
systems.

The basic setting, as used in \cite{VilelaJMP}, is the following:

In $\mathbb{R}^{4}$a sequence of hypercubic lattices is constructed in such
a way that any plaquette of edge size $\frac{a}{2^{k}}$ $\left(
k=0,1,2,\cdots \right) $\ is a refinement of a plaquette of edge $\frac{a}{%
2^{k-1}}$ (meaning that all vertices of the $\frac{a}{2^{k-1}}$ plaquette
are also vertices in the $\frac{a}{2^{k}}$ plaquette). The refinement is
made one-plaquette-at-a-time. Notice however that, when one plaquette of
edge $\frac{a}{2^{k-1}}$ is converted into four plaquettes of edge $\frac{a}{%
2^{k}}$, $2\times \left( d-2\right) $ new plaquettes of edge $\frac{a}{%
2^{k-1}}$, orthogonal to the refined plaquette, are also added to the
lattice. The additional plaquettes connect the new vertices of the refined $%
\frac{a}{2^{k}}$ plaquette to the middle points of $\frac{a}{2^{k-1}}$
plaquettes. Successive application of this process to all still unrefined $%
\frac{a}{2^{k-1}}$ plaquettes finally yields a full hypercubic $\frac{a}{%
2^{k}}$ lattice. See Fig.1 for a $3-$dimensional projection of the process,
where two of the additional four (in $\mathbb{R}^{4}$) plaquettes are shown,
attached to the points $A,B,C$ and $D$. This one-plaquette-at-a-time
construction is used to check the consistency condition (see Section 2).

Finite volume hypercubes $\Gamma $ in these lattices form a directed set $%
\left\{ \Gamma ,\succ \right\} $ under the inclusion relation $\succ $. $%
\Gamma \succ \Gamma ^{\prime }$ meaning that all edges and vertices in $%
\Gamma ^{\prime }$ are contained in $\Gamma $, the inclusion relation
satisfying%
\begin{eqnarray}
\Gamma &\succ &\Gamma  \notag \\
\Gamma &\succ &\Gamma ^{\prime }\text{ and }\Gamma ^{\prime }\succ \Gamma
\Longrightarrow \Gamma =\Gamma ^{\prime }  \notag \\
\Gamma &\succ &\Gamma ^{\prime }\text{ and }\Gamma ^{\prime }\succ \Gamma
^{\prime \prime }\Longrightarrow \Gamma \succ \Gamma ^{\prime \prime }
\label{I.1}
\end{eqnarray}%
Well-ordering of the directed set is insured by doing each new refinement in
the previously refined lattice. After each complete refinement of a finite
volume hypercube (from $\frac{a}{2^{k-1}}$ to $\frac{a}{2^{k}}$ size), the
sequence is expanded to include larger and larger volume hypercubes which
are likewise refined, etc..

\begin{figure}[htb]
\centering
\includegraphics[width=0.8\textwidth]{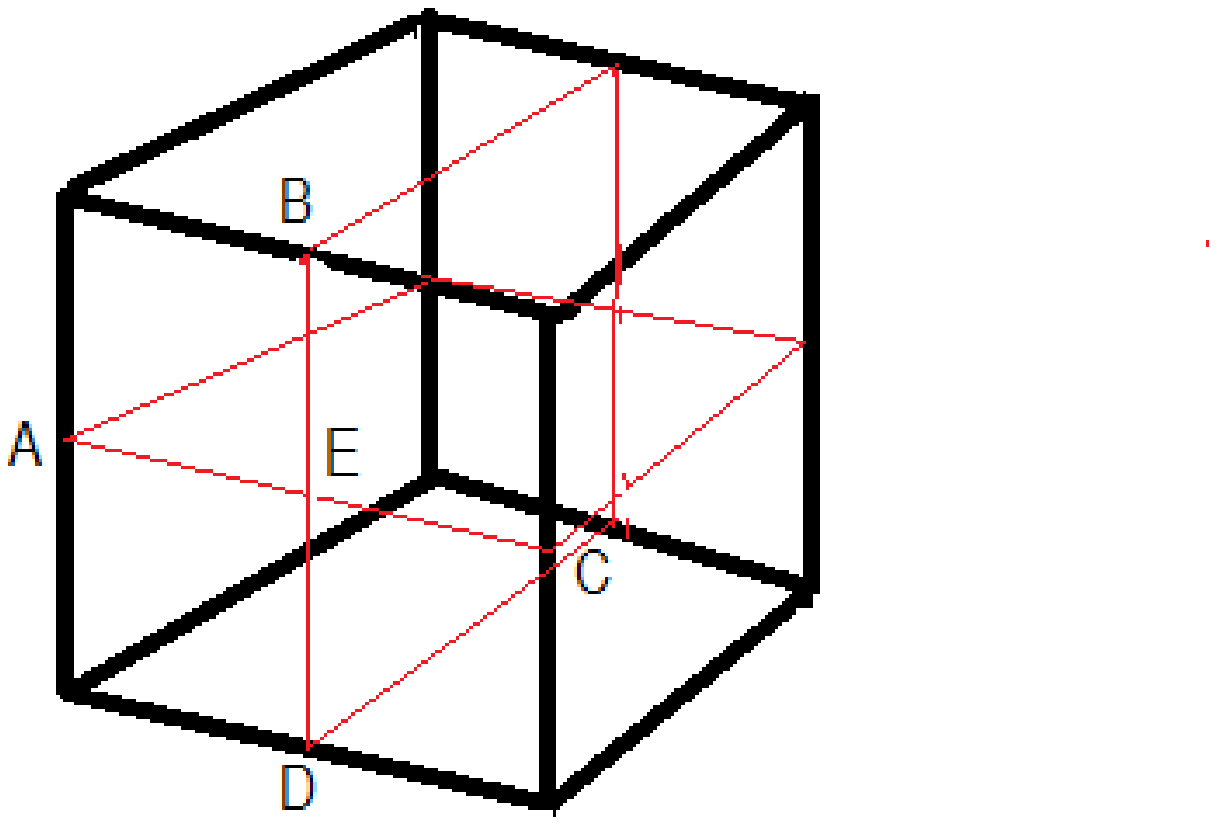}
\caption{Partial 3-dimensional projection of the one-plaquette-at-a-time
refinement process}
\end{figure}

Let $\mathbb{G}$ be a compact group and $x_{0}$ a point that does not belong
to any lattice point of the directed family. Assuming an analytic
parametrization of each edge, associate to each edge $l$ a $x_{0}$-based
loop and for each generalized connection $A$ consider the holonomy $%
h_{l}\left( A\right) $ associated to this loop. For definiteness each edge
is considered to be oriented along the coordinates positive direction and
the set of edges of the lattice $\Gamma $ is denoted $E\left( \Gamma \right) 
$. The set $\mathcal{A}_{\Gamma }$ of generalized connections for the
lattice hypercube $\Gamma $ is the set of homomorphisms $\mathcal{A}_{\Gamma
}=Hom\left( E\left( \Gamma \right) ,G\right) \sim G^{\#E\left( \Gamma
\right) }$, obtained by associating to each edge the holonomies $h_{l}\left(
\cdot \right) $ on the $x_{0}$-based loops that are associated to that edge. 
$G^{\#E\left( \Gamma \right) }$ is a product group, $\#E\left( \Gamma
\right) $ being the number of edges. The orientation of each $h_{l}\left(
\cdot \right) $ associated to an edge is the one compatible with the
orientation above defined for the edge. The set of gauge-independent
generalized connections $\mathcal{A}_{\Gamma }/Ad$ is obtained factoring by
the adjoint representation at $x_{0}$, $\mathcal{A}_{\Gamma }/Ad$ $\sim
G^{\#E\left( \Gamma \right) }/Ad$. However because, for gauge independent
functions, integration in $\mathcal{A}_{\Gamma }$ coincides with integration
in $\mathcal{A}_{\Gamma }/Ad$ , for simplicity, from now on one uses only $%
\mathcal{A}_{\Gamma }$. Finally one considers the projective limit $\mathcal{%
A}=\underset{\longleftarrow }{\lim }\mathcal{A}_{\Gamma }$ of the family 
\begin{equation}
\left\{ \mathcal{A}_{\Gamma },\pi _{\Gamma \Gamma ^{\prime }}:\Gamma
^{\prime }\succ \Gamma \right\}   \label{I.2}
\end{equation}%
$\pi _{\Gamma \Gamma ^{\prime }}$ and $\pi _{\Gamma }$ denoting the
surjective projections $\mathcal{A}_{\Gamma ^{\prime }}\longrightarrow 
\mathcal{A}_{\Gamma }$ and $\mathcal{A}\longrightarrow \mathcal{A}_{\Gamma }$%
.

The projective limit of the family $\left\{ \mathcal{A}_{\Gamma },\pi
_{\Gamma \Gamma ^{\prime }}\right\} $ is the subset $\mathcal{A}$ of the
Cartesian product $\underset{\Gamma }{\prod }\mathcal{A}_{\Gamma }$ defined
by 
\begin{equation}
\mathcal{A}=\left\{ a\in \underset{\Gamma }{\prod }\mathcal{A}_{\Gamma
}:\Gamma ^{\prime }\succ \Gamma \Longrightarrow \pi _{\Gamma \Gamma ^{\prime
}}a_{\Gamma ^{^{\prime }}}=a_{\Gamma }\right\}  \label{I.3}
\end{equation}%
with $a_{\Gamma ^{^{\prime }}}\in \mathcal{A}_{\Gamma ^{^{\prime }}}$, $%
a_{\Gamma }\in \mathcal{A}_{\Gamma }$ and the projective topology in $%
\mathcal{A}$ being the coarsest topology for which each $\pi _{\Gamma }$
mapping is continuous.

For a compact group $\mathbb{G}$, each $\mathcal{A}_{\Gamma }$ is a compact
Hausdorff space. Therefore $\mathcal{A}$ is also a compact Hausdorff space.
In each $\mathcal{A}_{\Gamma }$ one has a natural (Haar) normalized product
measure $\nu _{\Gamma }=\mu _{H}^{\#E\left( \Gamma \right) }$, $\mu _{H}$
being the normalized Haar measure in $\mathbb{G}$. Then, according to a
theorem of Prokhorov, as generalized by Kisynski \cite{Kisynski} \cite%
{Maurin}, if the following condition 
\begin{equation}
\nu _{\Gamma ^{\prime }}\left( \pi _{\Gamma \Gamma ^{\prime }}^{-1}\left(
B\right) \right) =\nu _{\Gamma }\left( B\right)   \label{I.4}
\end{equation}%
is satisfied for every $\Gamma ^{\prime }\succ \Gamma $ and every Borel set $%
B$ in $\mathcal{A}_{\Gamma }$, there is a unique measure $\nu $ in $\mathcal{%
A}$ such that $\nu \left( \pi _{\Gamma }^{-1}\left( B\right) \right) =\nu
_{\Gamma }\left( B\right) $ for every $\Gamma $. In this way a sequence of
measures is obtained that give the same weight to the sequence $\cdots
,B,\pi _{\Gamma \Gamma ^{\prime }}^{-1}\left( B\right) ,\cdots $ of  Borel
sets. In particular a continuum limit measure is obtained that is consistent
with the measures at each intermediate step of the lattice refinement.  

\section{The measure}

As stated before, the essential step in the construction of the measure in
the projective limit is the fulfilling of the consistency condition (\ref%
{I.4}). One considers, on the finite-dimensional spaces $\mathcal{A}_{\Gamma
}\sim G^{\#E\left( \Gamma \right) }$, measures that are absolutely
continuous with respect to the Haar measure 
\begin{equation}
d\mu _{\mathcal{A}_{\Gamma }}=p\left( \mathcal{A}_{\Gamma }\right) \left(
d\mu _{H}\right) ^{\#E\left( \Gamma \right) }  \label{II.1}
\end{equation}%
$p\left( \mathcal{A}_{\Gamma }\right) $ being a continuous function in $%
\mathcal{A}_{\Gamma }$ with the following two simplifying assumptions:

$\bullet $ $p\left( \mathcal{A}_{\Gamma }\right) $ is a product of plaquette
functions 
\begin{equation}
p\left( \mathcal{A}_{\Gamma }\right) =p\left( U_{\square _{1}}\right)
p\left( U_{\square _{2}}\right) \cdots p\left( U_{\square _{n}}\right)
\label{II.2}
\end{equation}%
with $U_{\square }\left( A_{\Gamma }\right) =h_{1}h_{2}h_{3}^{-1}h_{4}^{-1}$%
, $h_{1}$ to $h_{4}$ being the holonomies of the $x_{0}-$ based loops
associated to the edges of the plaquette, the orientation of the plaquette
being uniquely defined by the positive orientation of the edge to which it
is associated.

$\bullet $ $p\left( \cdot \right) $ is a central function, $p\left(
xy\right) =p\left( yx\right) $ or, equivalently $p\left( y^{-1}xy\right)
=p\left( x\right) $ with $x,y\in \mathbb{G}$.

\ \ \ 

Let $p^{\prime },p^{\prime \prime }$ and $p$ be the density functions
associated respectively to the square plaquette with edges of size $\frac{a}{%
2^{k}}$, to the rectangular plaquette with edges of size $\frac{a}{2^{k}}$
and $\frac{a}{2^{k-1}}$ and, finally, to the square plaquette with edges of
size $\frac{a}{2^{k-1}}$. Then

\textbf{Proposition 1 }\cite{VilelaJMP} \textit{A measure on the projective
limit }$\mathcal{A}=\underset{\longleftarrow }{\lim }\mathcal{A}_{\Gamma }$%
\textit{\ exists if a sequence of functions is found satisfying }%
\begin{eqnarray}
\int p^{\prime }\left( G_{i}X\right) p^{\prime }\left( X^{-1}G_{j}\right)
d\mu _{H}\left( X\right) &\sim &p^{\prime \prime }\left( G_{i}G_{j}\right) 
\notag \\
\int p^{\prime \prime }\left( G_{i}X\right) p^{\prime \prime }\left(
X^{-1}G_{j}\right) d\mu _{H}\left( X\right) &\sim &p\left( G_{i}G_{j}\right)
\label{II.3}
\end{eqnarray}%
\textit{for plaquette subdivisions of all sizes.}

\textbf{Proof:} In the directed set $\left\{ \Gamma ,\succ \right\} $
consider two elements $\Gamma $ and $\Gamma ^{^{\prime }}$ which differ only
in subdivision of a single plaquette from $\frac{a}{2^{k-1}}$ to $\frac{a}{%
2^{k}}$ size (see Fig.2) plus the additional $\frac{a}{2^{k-1}}$ plaquettes
(based on the middle points A, B, C and D) as explained in the introduction.

\begin{figure}[tbh]
\centering
\includegraphics[width=0.8\textwidth]{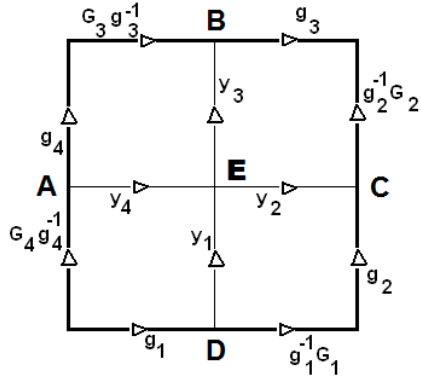}
\caption{Subdivision of one plaquette}
\end{figure}
To each edge one associates as many $x_{0}-$based loops as the number of
independent plaquettes that share that edge. For example to the edge
connecting the points A and C in Fig.2 there are four (in $\mathbb{R}^{4}$)
associated loops, two associated to the edges A-E and E-C, and two others
associated to the full edge A-C corresponding to the additional plaquettes
of size $\frac{1}{2^{k-1}}$. One associates the central function $p^{\prime
} $ to the first two loops and $p$ to the others. Notice that it is quite
consistent to associate more than one independent loop to each edge. The
integration is over the loops, not the edges.

Finally, the consistency condition (\ref{I.4}) requires that%
\begin{equation}
\begin{array}{l}
\frac{1}{Z^{^{\prime }}}\int p^{\prime }\left(
g_{1}^{-1}G_{1}g_{2}y_{2}^{-1}y_{1}^{-1}\right) p^{\prime }\left(
y_{2}g_{2}^{-1}G_{2}g_{3}^{-1}y_{3}^{-1}\right) p^{\prime }\left(
y_{4}y_{3}g_{3}G_{3}^{-1}g_{4}^{-1}\right) \\ 
p^{\prime }\left( g_{1}y_{1}y_{4}^{-1}g_{4}G\right) \prod_{i=1}^{4}d\mu
_{H}\left( g_{i}\right) d\mu _{H}\left( y_{i}\right) d\mu _{H}\left(
G_{i}\right) \prod_{k=1}^{2(d-2)}\left\{ p\left(
G_{1}^{(k)}G_{2}^{(k)}G_{3}^{(k)-1}G_{4}^{(k)-1}\right) \right. \\ 
\left. \prod_{j=1}^{4}d\mu _{H}\left( G_{j}^{(k)}\right) \right\} =\frac{1}{Z%
}\int p\left( G_{1}G_{2}G_{3}^{-1}G_{4}^{-1}\right) \prod_{i=1}^{4}d\mu
_{H}\left( G_{i}\right) .%
\end{array}
\label{II.4}
\end{equation}%
The last two factors in the left hand side%
\begin{equation}
\prod_{k=1}^{2(d-2)}\left\{ p\left(
G_{1}^{(k)}G_{2}^{(k)}G_{3}^{(k)-1}G_{4}^{(k)-1}\right) \prod_{j=1}^{4}d\mu
_{H}\left( G_{j}^{(k)}\right) \right\}  \label{II.4a}
\end{equation}
concern the integration over the additional $\frac{a}{2^{k}}$ plaquettes,
the density function $p$ used for these plaquettes being the one
corresponding to edges of size $\frac{a}{2^{k-1}}$ . $Z$ and $Z^{^{\prime }}$
are numerical constants related to normalization of the density functions.

Using centrality of $p^{\prime }$, redefining 
\begin{equation}
g_{1}y_{1}=X_{1},\qquad g_{2}y_{2}^{-1}=X_{2},\qquad
y_{3}g_{3}=X_{3}^{-1},\qquad y_{4}^{-1}g_{4}=X_{4}^{-1}  \label{II.5}
\end{equation}%
and using invariance of the normalized Haar measure, one may integrate over $%
y_{1},y_{2},y_{3},y_{4}$ and $G_{k}$, obtaining for the left hand side of (%
\ref{II.4}) with the exclusion of the terms in (\ref{II.4a})%
\begin{equation*}
\frac{1}{Z^{^{\prime }}}\int p^{\prime }\left( X_{1}^{-1}G_{1}X_{2}\right)
p^{\prime }\left( X_{2}^{-1}G_{2}X_{3}\right) p^{\prime }\left(
X_{3}^{-1}G_{3}^{-1}X_{4}\right) p^{\prime }\left(
X_{4}^{-1}G_{4}^{-1}X_{1}\right) \prod_{i=1}^{4}d\mu _{H}\left( X_{i}\right)
d\mu _{H}\left( G_{i}\right)
\end{equation*}%
Therefore if there is a sequence of central functions $p^{\prime },p^{\prime
\prime },p$ satisfying the proportionality relations 
\begin{eqnarray}
\int p^{\prime }\left( G_{i}X\right) p^{\prime }\left( X^{-1}G_{j}\right)
d\mu _{H}\left( X\right) &\sim &p^{\prime \prime }\left( G_{i}G_{j}\right) 
\notag \\
\int p^{\prime \prime }\left( G_{i}X\right) p^{\prime \prime }\left(
X^{-1}G_{j}\right) d\mu _{H}\left( X\right) &\sim &p\left( G_{i}G_{j}\right)
\label{II.6}
\end{eqnarray}%
the consistency condition (\ref{II.4}) would be satisfied, because the terms
in (\ref{II.4a}) dealing with integration over different loops may be
absorbed in the proportionality constant of the measure normalization. The
same procedure is then applied to all still unrefined plaquettes, meaning
that a measure would exist in the projective limit, because all elements in
the directed set $\left\{ \Gamma ,\succ \right\} $ may be reached by this
method. $\square $

With the conditions (\ref{II.3}) and the above construction the consistency
condition is satisfied by means of the pairwise convolutions of the $p$
functions for arbitrary dimensions. Notice also that by this refinement
method all plaquettes of a full $\frac{1}{2^{k}}$ lattice are obtained. For
example edges whose endpoints are at the center of a square of the coarser
lattice are obtained when one of the additional plaquettes of the above
process is also subdivided. Therefore, using a measure that satisfies the
condition (\ref{I.4}) one is sure that, in the continuum limit, a measure is
obtained that is consistent with the physical premises used to postulate a
measure for finite lattice spacing. This is an important feature, not only
for rigorous analytical developments, but also for the consistency of
numerical calculations at successively smaller lattice spacings.

If $p\left( U_{\square }\right) $ is a constant, $d\mu _{\mathcal{A}_{\Gamma
}}$ is factorizable and the consistency condition is trivially satisfied. $%
d\mu _{\mathcal{A}_{\Gamma }}$ would be the Ashtekar-Lewandowski measure for
generalized connections \cite{Ashtekar2} \cite{Ashtekar3}. A nontrivial
solution that satisfies the consistency condition (\ref{II.4}) is the choice
of $p\left( U_{\square }\right) $ as the heat kernel 
\begin{equation}
K\left( g,\beta \right) =\sum_{\lambda \in \Lambda ^{+}}d_{\lambda
}e^{-c\left( \lambda \right) \beta }\chi _{\lambda }\left( g\right)
\label{II.7}
\end{equation}%
with%
\begin{eqnarray}
\beta &\rightarrow &\beta ^{\prime }=\frac{\beta }{4}  \notag \\
\beta &\rightarrow &\beta ^{\prime \prime }=\frac{\beta }{2}  \label{II.8}
\end{eqnarray}%
$\beta ^{\prime \prime },\beta ^{\prime }$ and $\beta $ being the constants
associated to $p^{\prime \prime },p^{\prime }$ and $p$. In (\ref{II.7}), $%
g\in \mathbb{G}$, $\beta \in \mathbb{R}^{+}$, $\Lambda ^{+}$ is the set of
highest weights, $d_{\lambda }$ and $\chi _{\lambda }\left( \cdot \right) $
the dimension and character of the $\lambda -$representation and $c\left(
\lambda \right) $ the spectrum of the Laplacian $\Delta
_{G}:=\sum_{i=1}^{n}\chi _{i}^{2}$, $\left\{ \chi _{i}\right\} $ being a
basis for the Lie algebra of $\mathbb{G}$.

Finally, one writes for the measure on the lattice $\Gamma $%
\begin{equation}
d\mu _{\mathcal{A}_{\Gamma }}=\frac{1}{Z_{\Gamma }}\prod_{edges}d\mu
_{H}(g_{l})\prod_{plaquettes}\sum_{\lambda \in \Lambda ^{+}}d_{\lambda
}e^{-c\left( \lambda \right) \beta }\chi _{\lambda }\left( g_{p}\right)
\label{II.9}
\end{equation}%
and the consistency condition (\ref{I.4}) being satisfied, the
Prokhorov-Kisynski theorem \cite{Kisynski} insures that a measure is also
defined on the projective limit lattice, that is, on the projective limit
generalized connections $\mathcal{A}$.

This measure has the required naive continuum limit, both for abelian and
non-abelian theories (see \cite{VilelaJMP}). Furthermore by defining
infinite-dimensional test functionals and distributions, a projective
triplet was constructed which provides a framework to develop an
infinite-dimensional calculus over the hypercubical lattice. In particular,
this step is necessary to give a meaning to the density $p\left( \mathcal{A}%
_{\Gamma }\right) $ in the $\beta \rightarrow 0$ limit, where $p\left( 
\mathcal{A}_{\Gamma }\right) $ would no longer be a continuous function.
Thus $p\left( \mathcal{A}_{\Gamma }\right) $, a density that multiplies the
Ashtekar-Lewandowski measure \cite{Ashtekar2} \cite{Ashtekar3} \cite%
{Fleischhack3}, gains a distributional meaning in the framework of the
projective triplet.

A theory being completely determined whenever its measure is specified, the
construction in \cite{VilelaJMP} provides a rigorous specification of a
projective limit gauge field theory over a compact group. Some of the
consequences of this specification were already discussed in \cite{VilelaJMP}%
. Here one analyses the nature of the mass gap which follows from the
measure specification.

\section{The mass gap}

The experimental phenomenology of subnuclear physics provides evidence for
the short range of strong interactions. Therefore, if unbroken non-abelian
Yang-Mills is the theory of strong interactions, the Hamiltonian, associated
to its measure, should have a positive mass gap. This important physical
question has been addressed in different ways by several authors. An
interesting research approach \cite{Laufer1} \cite{Laufer2} considers the
Riemannian geometry of the (lattice) gauge-orbit space to compute the Ricci
curvature. The basic inspiration for this approach is the Bochner-Lichn\'{e}%
rowicz \cite{Bochner} \cite{Lichne} inequality which states that if the
Ricci curvature is bounded from below, then so is the first non-zero
eigenvalue of the Laplace-Beltrami operator. The Laplace-Beltrami operator
differs from the Yang-Mills Hamiltonian in that it lacks the chromo-magnetic
term, but the hope is that in the relevant physical limit the
chromo-electric term dominates the bound. An alternative possibility would
be to generalize the Bochner-Lichn\'{e}rowicz inequality.

Other approaches are based on attempts to solve the Dyson-Schwinger equation
(see for example \cite{SD1} \cite{SD2} \cite{SD3}) on a set of exact
solutions to the classical Yang-Mills theory \cite{Frasca} or on the
ellipticity of the energy operator of cut-off Yang-Mills \cite{Dynin1} \cite%
{Dynin2}.

Once a consistent Euclidean measure is obtained, the nature of the mass gap
may be found either by computing the distance dependence of the correlation
of two local operators or from the lower bound of the spectrum in the
corresponding Hamiltonian theory. Here the Hamiltonian approach will be
used, using the fact that the Hamiltonian may be obtained from the knowledge
of the ground state functional and the ground state functional may be
obtained from the measure \cite{Albeverio1} \cite{Hiroshima} \cite{Ludwig1} 
\cite{Ludwig2} \cite{Ludwig3} \cite{Vilela3}.

By inserting a complete set of energy states on the Euclidean path integral
and computing the integral pinned down at $t=0$ to a fixed configuration the
corresponding ground state may be obtained. This goes back to the work of
Donsker and Kac \cite{Donsker} \cite{Kac} and has been used and proved
before in several contexts \cite{Rossi} \cite{Fradkin}. At least for
finite-dimensional quantum systems this provides a robust estimation of the
ground state.

One of the axis directions in the lattice is chosen as the time direction.
Denote by $\theta \left( 0\right) $ the configuration of the system at time
zero. Then, recalling that at each step in the projective limit construction
one has a finite-dimensional system, the ground state wave functional $\Psi
_{0}\left( \theta \left( 0\right) \right) $ at the particular configuration $%
\theta \left( 0\right) $ may be written as%
\begin{eqnarray}
\left\vert \Psi _{0}\left( \theta \left( 0\right) \right) \right\vert ^{2}
&=&\int d\theta \Psi _{0}^{\ast }\left( \theta \right) \delta \left( \theta
-\theta \left( 0\right) \right) \Psi _{0}\left( \theta \right)  \notag \\
&=&\int d\mu _{\mathcal{A}}\left( \theta \right) \delta \left( \theta
-\theta \left( 0\right) \right)  \label{GS1}
\end{eqnarray}%
where $\mu _{\mathcal{A}}\left( \theta \right) $ is the Euclidean measure
and the integration is all variable configurations which in the time-zero
slice coincide with $\theta \left( 0\right) $. For the lattices considered
in this paper $\theta $ and $\theta \left( 0\right) $ stand respectively for
the set of group configurations in the $x_{0}-$based loops associated to the
edges and for the set of group configurations in the time-zero slice, namely 
$\theta _{j}^{\alpha }\left( l\right) $'s will be the Lie algebra
coordinates of the group elements $\exp \left( i\theta _{j}^{\alpha }\left(
l\right) \tau _{\alpha }\right) $, see Eq. (\ref{GS13}).

The ground state in (\ref{GS1}) may be used to develop the usual Hamiltonian
approach to lattice theory, for which one uses notations similar to those of
Chapter 15 in Ref.\cite{Creutz}, the main difference being that instead of
constructing the Kogut-Susskind Hamiltonian from the Wilson action, one uses
the ground state obtained from the measure.

The squared wave-function in (\ref{GS1}) is the density of a \textit{ground
state measure} \cite{Albeverio1} \cite{Hiroshima}. Associated to this ground
state measure, there is a stochastic process for which the measure is
invariant. The canonical way \cite{Ludwig1} \cite{Ludwig2} \cite{Ludwig3} 
\cite{Vilela3} to construct the elliptic operator generator of the process is%
\begin{equation}
H_{g}^{\prime }=\frac{g^{2}\left( \beta \right) }{2\beta }\sum_{l,j,\alpha
}\left\{ -\frac{\partial }{\partial \theta _{j}^{\alpha }\left( l\right) }%
+L_{j}^{\alpha }\left( l\right) \right\} \left\{ \frac{\partial }{\partial
\theta _{j}^{\alpha }\left( l\right) }+L_{j}^{\alpha }\left( l\right)
\right\}  \label{GS2}
\end{equation}%
with%
\begin{equation}
L_{j}^{\alpha }\left( l\right) =-\frac{1}{\Psi _{0}}\frac{\partial \Psi _{0}%
}{\partial \theta _{j}^{\alpha }\left( l\right) }  \label{GS3}
\end{equation}%
With the unitary transformation $H_{g}^{\prime }\rightarrow H_{g}=\Psi
_{0}H_{g}^{\prime }\Psi _{0}^{-1}$ the operator in (\ref{GS2}) would have
the familiar form of Laplacian plus potential. The $\theta _{j}^{\alpha
}\left( l\right) $'s are the Lie algebra coordinates of the group element $%
\exp \left( i\theta _{j}^{\alpha }\left( l\right) \tau _{\alpha }\right) $
at each $x_{0}-$based loop associated to the edge $l$ of the time-zero slice
of the lattice, the sum being over edges $\left( l\right) $, lattice
dimensions $\left( j\right) $ and Lie algebra generators $\left( \alpha
\right) $. $g\left( \beta \right) $ is a coupling constant to be adjusted
consistently to obtain the continuum limit, to be discussed later. Recall
that from (\ref{II.8}) $\beta \rightarrow 0$ as the length of the lattice
edges ($\frac{a}{2^{k}}$) goes to zero. Eq.(\ref{GS2}) implies that the
ground state energy $E_{0}$ is adjusted to zero

In this way a Hamiltonian and a Hilbert space may be constructed from the
Euclidean measure and estimations of the principal eigenvalue may be
obtained from the theory of small random perturbations of dynamical systems 
\cite{Friedman} \cite{Freidlin1}. These steps are briefly summarized below
and then applied to the lattices of the projective family.

Explicit computation of the integral in (\ref{GS1}) is, in general, not
easy. However, to study the nature of the mass gap a full calculation of the
ground state functional is not required. The interpretation of elliptic
operators as generators of a diffusion process \cite{Friedman} \cite%
{Freidlin1} may be used and, in the limit of small $\beta $, also the theory
of small perturbations of dynamical systems \cite{Freidlin2} \cite{Freidlin3}%
.

For simplicity, Eq.(\ref{GS2}) applies to steps of the projective limit when
the same uniform $\beta $ exists throughout the lattice. For intermediate
steps of the refinement process, a slightly more complex definition would
apply. This however will not change the main conclusions. At each step of
the projective limit construction one deals with a finite dimensional
quantum system. For the Hamiltonian Then, absence of zeros in the ground
state allows the unitary transformation $H_{g}^{\prime }=\Psi
_{0}^{-1}H_{g}\Psi _{0}$, the ground state $\Psi _{0}$ is the unit function,
the corresponding states of $H_{g}$ being multiplied by $\Psi _{0}^{-1}$%
\begin{equation}
-\beta H_{g}^{\prime }=\frac{g^{2}\left( \beta \right) }{2}\sum_{l,j,\alpha }%
\frac{\partial }{\partial \theta _{j}^{\alpha }\left( l\right) }\frac{%
\partial }{\partial \theta _{j}^{\alpha }\left( l\right) }+\sum_{l,j,\alpha
}b_{j}^{\alpha }\left( l\right) \frac{\partial }{\partial \theta
_{j}^{\alpha }\left( l\right) }  \label{GS4}
\end{equation}%
with%
\begin{equation}
b_{j}^{\alpha }\left( l\right) =-g^{2}\left( \beta \right) L_{j}^{\alpha
}\left( l\right) =\frac{g^{2}\left( \beta \right) }{2\Psi _{0}^{2}}\frac{%
\partial \ln \Psi _{0}^{2}}{\partial \theta _{j}^{\alpha }\left( l\right) }
\label{GS5}
\end{equation}%
The second-order elliptic operator in (\ref{GS4}) is the generator of the
diffusion process%
\begin{equation}
d\theta _{j}^{\alpha }\left( l\right) =b_{j}^{\alpha }\left( l\right)
dt+g\left( \beta \right) dW_{j}^{\alpha }\left( l\right)  \label{GS6}
\end{equation}%
with drift $b_{j}^{\alpha }\left( l\right) $ and diffusion coefficient $%
g\left( \beta \right) $. $\left\{ W_{j}^{\alpha }\left( l\right) \right\} $
is a set of independent Brownian motions. $\Psi _{0}^{2}$ is the invariant
measure of this process. The question of existence of a mass gap for the
Hamiltonian $H_{g}^{\prime }$ is closely related to the principal eigenvalue
of the Dirichlet problem%
\begin{eqnarray}
\beta H_{g}^{\prime }u &=&\lambda u\hspace{1.5cm}\text{in }D  \notag \\
u &=&0\hspace{1.5cm}\text{in }\partial D  \label{GS6a}
\end{eqnarray}%
$D$ being a bounded domain (in the space of the variables) and $\partial D$
its boundary. The principal eigenvalue $\lambda _{1}$, that is, the smallest
positive eigenvalue of $\beta H_{g}^{\prime }$ has a stochastic
representation \cite{Freidlin3} \cite{Khasminskii}%
\begin{equation}
\lambda _{1}=\sup \left\{ \lambda \geq 0;\sup_{\theta \in D}\mathbb{E}%
_{\theta }e^{\lambda \tau }<\infty \right\}  \label{GS6b}
\end{equation}%
$\mathbb{E}_{\theta }$ denoting the expectation value for the process
started from the $\theta $ configuration and $\tau $ the time of first exit
from the domain $D$. The validity of this result hinges on the following
condition

\textbf{(C1)} The drift $b$ and the diffusion matrix coefficient $\sigma $ ($%
g\left( a\right) \delta _{ij}$ in this case) must be uniformly Lipschitz
continuous with exponent $0<\alpha \leq 1$ and $\sigma $ positive definite.

(\ref{GS6b}) is a powerful result which may be used to compute by numerical
means the principal eigenvalue for arbitrary values of $g$ \footnote{%
See for example Ref. \cite{Eleuterio}}. However, a particularly useful
situation is the small noise (small $g$ limit). That the small noise limit
corresponds to the continuum limit of the lattice theory follows from a
consistency argument. Under suitable conditions, to be discussed below, the
small noise limit of the lowest nonzero eigenvalue (the mass gap) of the
operator $\beta H^{\prime }$ is%
\begin{equation}
\beta m\sim \exp \left( -\frac{V}{g^{2}\left( \beta \right) }\right)
\label{GS7}
\end{equation}%
where $V$ is the value of a functional. Hence, for the physical mass gap $m$
to remain fixed when $\beta \rightarrow 0$, it should also be $g\left( \beta
\right) \rightarrow 0$. Therefore the small noise limit is indeed the
continuum limit.

In the small noise limit the mass gap may be obtained from the
Wentzell-Freidlin estimates \cite{Freidlin2} \cite{Freidlin3}. Given a
bounded domain $D$ for the variables $\theta _{j}^{\alpha }\left( l\right) $
define the functional%
\begin{equation}
I_{t_{1},t_{2}}\left( \chi \right) =\frac{1}{2}\int_{t_{1}}^{t_{2}}\left( 
\frac{d\chi }{ds}-b\left( \chi \left( s\right) \right) \right) ^{2}ds
\label{GS8}
\end{equation}%
where $\chi \left( s\in \left[ t_{1},t_{2}\right] \right) $ is a path from
the configuration $\left\{ \theta \right\} $ to the boundary $\partial D$ of
the domain $D$. Then let%
\begin{equation}
I\left( t,\left\{ \theta \right\} ,\partial D\right) =\inf_{\chi
}I_{0,t}\left( \chi \right)  \label{GS9}
\end{equation}%
be the infimum over all continuous paths that starting from the
configuration $\left\{ \theta \right\} $ hit the boundary $\partial D$ in
time less than or equal to $t$. A path is said to be a \textit{neutral path}
if $I\left( t,\left\{ \theta \right\} ,\partial D\right) =0$.

The value of this functional is controlled by the nature of the
deterministic dynamical system%
\begin{equation}
\frac{d\theta _{j}^{\alpha }\left( l\right) }{dt}=b_{j}^{\alpha }\left(
l\right)  \label{GS10}
\end{equation}%
Assume the following additional condition to be fulfilled:

\textbf{(C2)} There are a number $r$ of $\omega -$limit sets $K_{i}$ of (\ref%
{GS10}) in the domain $D$, with all points in each set $K_{i}$ being
equivalent for the functional $I$, that is, $I\left( t,x,y\right) =0$ if
both $x,y\in K_{i}$ and $b\bullet \nu >0$, $\nu $ being the inward normal to 
$\partial D$.

Then \cite{Friedman} \cite{Freidlin3} with%
\begin{equation}
V_{i}=\inf I\left( t,x,\partial D\right) \hspace{1cm}\text{for }x\in K_{i}
\label{GS11}
\end{equation}%
and%
\begin{eqnarray*}
V_{\ast } &=&\max \left( V_{1},\cdots ,V_{r}\right) \\
V^{\ast } &=&\min \left( V_{1},\cdots ,V_{r}\right)
\end{eqnarray*}%
the lowest non-zero eigenvalue $\lambda _{1}$ satisfies%
\begin{equation*}
V_{\ast }\leq \lim_{g\rightarrow 0}\left( -g^{2}\ln \lambda _{1}\left(
g\right) \right) \leq V^{\ast }
\end{equation*}%
In particular if there is only one $V$%
\begin{equation}
\lambda _{1}\left( g\right) =\beta m\left( g\right) \asymp \exp \left( -%
\frac{V}{g^{2}\left( \beta \right) }\right)  \label{GS12}
\end{equation}%
the symbol $\asymp $ meaning logarithmic equivalence in the sense of large
deviation theory. If the drift is the gradient of a function, as in (\ref%
{GS5}), the quasi-potential $V$ is simply obtained from the difference of
the function at the $\omega -$limit set and the minimum at the boundary.

For details on the theory of small perturbations of dynamical systems as
applied to the small $\beta $ limit of lattice theory refer also to \cite%
{Vilela2} where this technique was applied to an approximate ground state
functional. Also \cite{Ludwig1} \cite{Ludwig2} \cite{Ludwig3} \cite{Vilela3}
provide details on how the ground state measure provides a complete
specification of quantum theories both for local and non-local potentials.
This theory developed for finite-dimensional systems follows earlier
developments of Coester, Haag and Araki \cite{Coester} \cite{Araki} in the
field theory context.

Now the existence of a mass gap associated to the Hamiltonian (\ref{GS4}),
obtained from the measure (\ref{II.9}) by (\ref{GS1}), hinges on checking
the above conditions \textbf{(C1)} and \textbf{(C2)}. Inserting (\ref{II.9})
into (\ref{GS1}) one obtains%
\begin{equation}
\left\vert \Psi _{0}\left( g_{l}\left( 0\right) \right) \right\vert
^{2}=\int \prod_{edges}d\mu _{H}(g_{l})\delta \left( g_{l}-g_{l}\left(
0\right) \right) \prod_{plaquettes}\sum_{\lambda \in \Lambda ^{+}}d_{\lambda
}e^{-c\left( \lambda \right) \beta }\chi _{\lambda }\left( g_{p}\right)
\label{GS13}
\end{equation}%
$g_{l}$ being the group element associated to the edge-associated loops and $%
g_{p}$ those associated to the ordered product of group elements around a
plaquette, $\left\vert \Psi _{0}\left( g_{l}\left( 0\right) \right)
\right\vert ^{2}$ being a function only of the group elements on the time
slice. For practical calculations one makes a global lattice gauge fixing in
(\ref{GS13}) but for the present considerations this is not important.

In (\ref{GS13}) the only free variables are the edge variables in the time
slice or, more precisely, the angles of the maximal torus of the group
element associated to the corresponding plaquettes. Smoothness of the heat
kernel implies that the Leibnitz rule for derivation under the integral can
be applied and the drift $b_{j}^{\alpha }\left( l\right) $ in (\ref{GS10})
is also a smooth function. Therefore condition \textbf{(C1)} is satisfied.
As for condition \textbf{(C2)} one knows that the heat kernel satisfies the
following two-sided Gaussian estimate%
\begin{equation}
\frac{1}{\left\vert B\left( e,\beta ^{\frac{1}{2}}\right) \right\vert }%
c_{1}\exp \left( \frac{-d^{2}\left( g\right) }{c_{2}\beta }\right) \leq
K\left( g,\beta \right) \leq \frac{1}{\left\vert B\left( e,\beta ^{\frac{1}{2%
}}\right) \right\vert }c_{3}\exp \left( \frac{-d^{2}\left( g\right) }{%
c_{4}\beta }\right)  \label{GS15}
\end{equation}%
$d\left( g\right) $ being the Carnot-Carath\'{e}odory distance of the group
element $g$ to the identity $e$ and $\left\vert B\left( e,\beta ^{\frac{1}{2}%
}\right) \right\vert $ is the volume of a ball of radius $\beta ^{\frac{1}{2}%
}$ centered at $e$ \cite{Saloff} \cite{Varopoulos}. The estimate (\ref{GS15}%
) holds if and only if

(A) the volume growth has the doubling property%
\begin{equation*}
\forall x\in \mathbb{G},\forall r>0,\left\vert B\left( x,2r\right)
\right\vert \leq c\left\vert B\left( x,r\right) \right\vert
\end{equation*}

(B) there is a constant $\gamma $ such that%
\begin{equation*}
\forall x\in \mathbb{G},\forall r>0,\int_{B\left( x,r\right) }\left\vert
f-Av_{B\left( x,r\right) }f\right\vert ^{2}dx\leq \gamma r^{2}\int_{B\left(
x,2r\right) }\left\vert \nabla f\right\vert ^{2}
\end{equation*}%
$Av_{B\left( x,r\right) }f$ being the average of $f$ over the ball $B\left(
x,r\right) $. In particular if $\mathbb{G}$ is unimodular (B) holds.

For a compact group (A) and (B) being satisfied, the two-sided estimate (\ref%
{GS15}) holds. Therefore the dynamical system (\ref{GS10}) has only one $%
\omega -$limit set, the group identity, and one is in the situation of Eq.(%
\ref{GS12}), $V$ being obtained from the difference of the heat kernel at
the identity and at the boundary of the domain. In conclusion:

\textbf{Proposition 2 }\textit{If }$G$\textit{\ is a compact group, the
Hamiltonian (\ref{GS6}) obtained from the heat-kernel measure has a positive
mass gap.}

The Wentzell-Freitlin results that are used to reach this result apply to a
Dirichlet problem with boundary. Therefore one is in fact considering some
bounded domain in the group space containing the identity, not necessarily
the full group space. This is probably consistent because for small $\beta $
the measure contributions are dominated by group elements close to the
identity (see the comparison with the na\"{\i}ve continuum limit in Ref.\cite%
{VilelaJMP}). However this is an issue that might deserve further
consideration.

The result is obtained for the Hamiltonians constructed from the Euclidean
measure constructed for each finite dimensional lattice in the directed set $%
\left\{ \Gamma ,\succ \right\} $. By itself, the result depends on the
nature of the central functions $p$ chosen in (\ref{II.7}), and not on the
consistency property and the existence of the projective limit measure.
However, what \textit{the specific form of the mass gap (\ref{GS12}) and the
consistency of the measure together imply is that there is a choice }$%
g\left( \beta \right) $\textit{\ that allows the construction of a continuum
limit theory (at }$\beta \rightarrow 0$\textit{) with a finite mass gap}.

This mass gap result, being based on the small random perturbations
(Wentzell-Freidlin (WF)) estimates, is of a strictly non-perturbative (NP)
nature. The WF estimates are in fact a tool of choice for NP reasoning
because they have at all orders an essential singularity on the coupling
(noise) constant.

The projective limit, being the subset of the direct product of all lattice
refinements that satisfies a consistent condition (Eq.(\ref{I.3})), it
describes a framework for all length scales, with a consistent measure down
to the vanishing lattice space limit. Therefore, this construction may then
be considered as a scaffold for the physical theory, which is embedded in
the projective limit structure by a Hamiltonian constructed from the measure
up to a (coupling) constant. Uniform physical results are obtained by the
choice at each length scale of the free parameter (the coupling constant).
In particular, to obtain a finite mass gap at all length scales, it is
indeed needed to make the coupling constant approach zero with the lattice
spacing. However, this weak coupling limit is fully non-perturbative because
based on an estimate with an essential singularity.

The existence of the projective limit\ measure, the projective triplet,
consistency with the required physical continuum limit as shown in \cite%
{VilelaJMP}, as well as the characterization of the nature of the mass gap
obtained here, might provide a consistent constructive definition of a
theory that might serve the physical purposes aimed at by the Yang-Mills
action. Of course, to scale up these results to a full understanding of QCD
the role of fermions as well as of the non-generic strata \cite{Vilela2b}
would be required. In particular to clarify the importance of these strata
for the structure of low-lying excitations.


\begin{thebibliography}{99}
\bibitem{VilelaJMP} R. Vilela Mendes; \textit{An infinite-dimensional
calculus for generalized connections in hypercubic lattices}, J. Math. Phys.
52 (2011) 052304.

\bibitem{Freidlin2} A. D. Wentzell and M. I. Freidlin; \textit{On small
random perturbations of dynamical systems}, Russian Math. Surveys 25 (1970)
1--55.

\bibitem{Freidlin3} M. I. Freidlin and A. D. Wentzell; \textit{Random
perturbations of dynamical systems}, Springer, Berlin 2012.

\bibitem{Kisynski} J. Kisynski;\textit{\ On the generation of tight measures}%
, Studia Math. 30 (1968) 141-151.

\bibitem{Maurin} K. Maurin; \textit{General eigenfunction expansions and
unitary representations of topological groups}, PWN - Polish Scient. Publ.,
Warszawa 1968.

\bibitem{Ashtekar2} A. Ashtekar and J. Lewandowski; \textit{Differential
geometry on the space of connections via graphs and projective limits}, J.
Geom. Phys. 17 (1995) 191-230.

\bibitem{Ashtekar3} A. Ashtekar and J. Lewandowski; \textit{Projective
techniques and functional integration for gauge theories}, J.Math. Phys. 36
(1995) 2170-2191.

\bibitem{Fleischhack3} C. Fleischhack; \textit{On the support of physical
measures in gauge theories}, arXiv:math-ph/0109030.

\bibitem{Laufer1} M. S. Laufer and P. Orland; \textit{The metric of
Yang-Mills orbit space on the lattice}, Phys.Rev. D88 (2013) 065018

\bibitem{Laufer2} M. S. Laufer; \textit{The Geometry of Lattice-Gauge-Orbit
Space}, Ph. D. Thesis The City University of New York, 2011.

\bibitem{Bochner} S. Bochner; \textit{Vector fields and Ricci curvature},
Bull. Amer. Math. Soc. 52 (1946) 776-797.

\bibitem{Lichne} A. Lichn\'{e}rowicz; \textit{G\'{e}ometrie des groupes de
transformations}, Dunod, Paris 1958.

\bibitem{SD1} R. Alkofer, A. Hauck, L. von Smekal; \textit{Infrared Behavior
of Gluon and Ghost Propagators in Landau Gauge QCD}, Physical Review Letters
79 (1997) 3591-3594.

\bibitem{SD2} V. Gogokhia; \textit{How to demonstrate a possible existence
of a mass gap in QCD}, arXiv:hep-th/0604095v4

\bibitem{SD3} B. Holdom; \textit{Soft asymptotics with mass gap}, Physics
Letters B 728 (2014) 467--471.

\bibitem{Frasca} M. Frasca; \textit{Exact solutions for classical Yang-Mills
fields}, \qquad arXiv:1409.2351

\bibitem{Dynin1} A. Dynin; \textit{Quantum Yang--Mills--Weyl dynamics in the
Schroedinger paradigm,} Russian Journal of Mathematical Physics 21 (2014)
169--188.

\bibitem{Dynin2} A. Dynin; \textit{On the Yang--Mills Mass Gap Problem},
Russian Journal of Mathematical Physics 21 (2014) 326--328.

\bibitem{Albeverio1} S. Albeverio, Yu. G. Kondratiev, R. A. Minlos and G. V.
Shchepa'uk; \textit{Ground state Euclidean measures for quantum lattice
systems on compact manifolds}, Rep. Math. Phys. 45 (2000) 419-429.

\bibitem{Hiroshima} F. Hiroshima; \textit{Ground state measure and its
applications}, RIMS pub. 1156, pp. 74-84, 2000.

\bibitem{Ludwig1} S. Albeverio, R. H\o egh-Krohn and L. Streit; \textit{%
Energy Forms, Hamiltonians, and Distorted Brownian Paths}, J. Math. Phys. 18
(1977) 907-917.

\bibitem{Ludwig2} S. Albeverio, R. H\o egh-Krohn and L. Streit; \textit{%
Regularization of Hamiltonians and Processes}, J. Math. Phys. 21 (1980)
1636-1642.

\bibitem{Ludwig3} L. Streit; \textit{Energy forms: Schroedinger theory,
processes}, Physics Reports 77 (1981) 363-375.

\bibitem{Vilela3} R. Vilela Mendes; \textit{Reconstruction of dynamics from
an eigenstate}, J. of Math. Phys. 27 (1986) 178-184.

\bibitem{Donsker} M. D. Donsker and M. Kac; \textit{Sampling method for
determining the lowest eigenvalue and the principal eigenfunction of Schr%
\"{o}dinger equation}, J. Res. Nat. Bur. Standards 44 (1950) 551-557.

\bibitem{Kac} M. Kac; \textit{On some connections between probability theory
and differential and integral equations}, in Proc. of the second Berkeley
Symposium on Math. Stat. and Probability, 189-215, Berbeley Press 1951.

\bibitem{Rossi} G. C. Rossi and M. Testa; \textit{Ground State Wave Function
from Euclidean Path Integral,} Annals of Physics 148 (1983) 144-167.

\bibitem{Fradkin} E. Fradkin; \textit{Wave functionals for field theories
and path integrals, }Nuclear Physics B389 (1993) 587-600.

\bibitem{Creutz} M. Creutz; \textit{Quarks, gluons and lattices}, Cambridge
U. P., Cambridge 1983.

\bibitem{Friedman} A. Friedman; \textit{Stochastic differential equations
and applications, vol. 2}, Academic Press, New York 1976.

\bibitem{Freidlin1} M. Freidlin; \textit{Markov processes and differential
equations: Asymptotic problems}, Birkh\"{a}user, Basel 1996.

\bibitem{Khasminskii} R. Z. Khas'minskii; \textit{On positive solutions of
the equation }$\mathfrak{U}$\textit{u + V \textperiodcentered\ u = 0},
Theory Probab. Appl. 4 (1959) 309--318.

\bibitem{Eleuterio} S. M. Eleut\'{e}rio and R. Vilela Mendes; \textit{%
Numerical predictions from a stochastic model for SU(2) lattice gauge fields}%
, Phys. Lett. B173 (1986) 332-336.

\bibitem{Vilela2} R. Vilela Mendes; \textit{Stochastic processes and the
non-perturbative structure of the QCD vacuum}, Z. Phys. C - Particles and
Fields 54 (1992)\ 273-281.

\bibitem{Coester} F. Coester and R. Haag, \textit{Representation of states
in a field theory with canonical variables}, Phys. Rev. 117 (1970) 1137-1145.

\bibitem{Araki} H. Araki, \textit{Hamiltonian formalism and the canonical
commutation relations in quantum field theory}, J. Math. Phys. 1 (1960)
492-504.

\bibitem{Saloff} L. Saloff-Coste; \textit{Aspects of Sobolev-type
inequalities}, Cambridge Lect. Notes 289, Cambridge Univ. Press, Cambridge
2002.

\bibitem{Varopoulos} N. Th. Varopoulos.L. Saloff-Coste and T. Coulhon; 
\textit{Analysis and geometry on groups}, Cambridge Tracts on Math. 100,
Cambridge Univ. Press, Cambridge 1992.

\bibitem{Vilela2b} R. Vilela Mendes; \textit{Stratification of the orbit
space in gauge theories. The role of nongeneric strata}, J. Phys. A: Math.
Gen. 37 (2004) 11485-11498.
\end{thebibliography}
\end{document}